\documentclass[a4paper,mathleft]{an}
\usepackage{graphicx}
\usepackage{epsfig}
\usepackage{times}
\begin{document}


\title{AsteroFLAG: first results from hare-and-hounds Exercise\,\#1}

\author{W.~J.~Chaplin\inst{1}\fnmsep\thanks{Corresponding author:
\email{w.j.chaplin@bham.ac.uk}\newline} \and T.~Appourchaux\inst{2}
\and T.~Arentoft\inst{3} \and J.~Ballot\inst{4} \and
J.~Christensen-Dalsgaard\inst{3} \and O.~L.~Creevey\inst{5,6} \and
Y.~Elsworth\inst{1} \and S.~T.~Fletcher\inst{7} \and
R.~A.~Garc\'ia\inst{8} \and G.~Houdek\inst{9} \and
S.~J.~Jim\'enez-Reyes\inst{6} \and H.~Kjeldsen\inst{3} \and
R.~New\inst{7} \and C.~R\'egulo\inst{10,6} \and D.~Salabert\inst{11}
\and T.~Sekii\inst{12} \and S.~G.~Sousa\inst{13,14} \and
T.~Toutain\inst{1} \and the rest of the asteroFLAG group\inst{15}}
\titlerunning{asteroFLAG hare and hounds} \authorrunning{W. J. Chaplin
et al.}

\institute{School of Physics and Astronomy, University of Birmingham,
Edgbaston, Birmingham, B15 2TT, UK \and Institut d'Astrophysique
Spatiale (IAS), Batiment 121, F-91405, Orsa Cedex, France \and
Department of Physics and Astronomy, University of Aarhus, DK-8000
Aarhus C, Denmark \and Max-Planck-Institut f\"ur Astrophysik,
Karl-Schwarzschild-Str. 1, Postfach 1317, 85741, Garching, Germany
\and High Altitude Observatory, National Center for Atmospheric
Research, Boulder, CO 80301, USA \and Instituto de Astrof\'\i sica de
Canarias, E-38200, La Laguna, Tenerife, Spain \and Faculty of Arts,
Computing, Engineering and Sciences, Sheffield Hallam University,
Sheffield S1 1WB, UK \and DAPNIA/CEA, CE Saclay, FR-91191
Gif-sur-Yvette Cedex, France \and Institute of Astronomy, University
of Cambridge, Cambridge CB3 0HA, UK \and Dpto. de Astrof\'isica,
Universidad de La Laguna, La Laguna, 38206, Tenerife, Spain \and
National Solar Observatory, 950 North Cherry Avenue, Tucson, AZ 85719,
USA \and National Astronomical Observatory of Japan, Mitaka, Tokyo,
181-8588, Japan \and Centro de Astrof\'isica Universidade do Porto,
4150-762 Porto, Portugal \and Departamento de Matem\'atica Aplicada,
Faculdade de Ci\^encias da Universidade do Porto, Portugal \and From a
further 4 institutes}

\received{20 Oct 2007}
\accepted{XX Xxx 200X}
\publonline{later}


\abstract{We report on initial results from the first phase of
Exercise\,\#1 of the asteroFLAG hare and hounds. The asteroFLAG group
is helping to prepare for the asteroseismology component of NASA's
Kepler mission, and the first phase of Exercise\,\#1 is concerned with
testing extraction of estimates of the large and small frequency
spacings of the low-degree p modes from Kepler-like artificial
data. These seismic frequency spacings will provide key input for
complementing the exoplanet search data.}

\maketitle

 \section{Introduction}
 \label{sec:intro}

With the recent launch of CoRoT (Baglin et al. 2006), the upcoming
launch of Kepler (Basri et al. 2005), and continuation of observations
by ground-based teams (Bedding \& Kjeldsen 2006, 2007) and MOST
(Matthews et al. 2007), we stand on the threshold of a critical
expansion of asteroseismology of Sun-like stars, the study of stellar
interiors by observation of their global acoustic modes of
oscillation. Sun-like oscillations give a very rich spectrum allowing
the internal structures and dynamics to be probed down into the
stellar cores to very high precision. Asteroseismic observations of
many stars will allow multiple-point tests of crucial aspects of
stellar evolution and dynamo theory.

The aims of the asteroFLAG collaboration are to help the community to
refine existing methods for analysis of the asteroseismic data on
Sun-like stars and to develop new ones.  The asteroFLAG group is
diverse.  There is expertise from those involved in the cutting-edge
ground-based asteroseismic observations; members of the CoRoT Data
Analysis Team (DAT) (see Appourchaux 2003; 2006a, b); members of the
Kepler Asteroseismology Science Consortium (KASC)
(Christensen-Dalsgaard et al. 2007); leading theoreticians on the
interior structures of stars and Sun-like oscillations; and those
involved in the analysis of the so-called ``Sun-as-a-star''
helioseismology data, for which the analysis methods have direct
application to the asteroseismic case (e.g., Chaplin et al. 2006).

The input data for probing stellar interiors are the mode parameters,
such as individual frequencies, frequency splittings, linewidths and
powers. For those stars where measurement of individual mode
parameters is difficult, the input data will be the likes of average
frequency spacings. Accurate mode parameter data are a vital
prerequisite for robust, accurate inference on the internal structures
of the stars.  Our objectives are to test aspects of the complete
analysis pipelines for stars, i.e., extraction of estimates of the
mode parameters from observations, through to procedures used to draw
inference on the fundamental stellar properties and the internal
structures.

Initially, asteroFLAG is concentrating on main-sequence stars and is
conducting the work within a \emph{hare-and-hounds} framework. Sets of
artificial asteroseismic data will be made by \emph{hares} in the
group, to simulate observations by different instruments (both ground
and space-based). Information for constructing the artificial data
comes from several sources, including full stellar evolutionary codes
and analytical descriptions of the stochastic excitation and damping
of the Sun-like oscillations. The artificial data are then being
analyzed by other members of the group, who are the \emph{hounds}.

AsteroFLAG is already involved in helping to prepare for the
asteroseismology component of NASA's Kepler mission. In this paper, we
report on the initial results of the first phase of Exercise\,\#1 of
the asteroFLAG hare and hounds, which is concerned with extraction of
the large and small frequency spacings of the p-mode spectra, from
Kepler-like observations of main-sequence stars.

 \section{Exercise\,\#1 of asteroFLAG hare and hounds: Observations by Kepler}
 \label{sec:HH1}

In addition to searching for Earth-like exoplanets (via the transit
method), NASA's Kepler mission will also provide an unprecedented
opportunity to study a wide range of stars by asteroseismology.  The
Kepler Asteroseismology Investigation (KAI) will be arranged around
the Kepler Asteroseismic Science Operations Centre (KASOC), which will
be based at the Danish AsteroSeismology Centre (DASC, Aarhus)
(Christensen-Dalsgaard et al. 2007). There is wide participation from
the international community in the Kepler Asteroseismology Science
Consortium (KASC).

The nominal mission lifetime is $\approx 4$ years. There will
potentially be a few hundred Sun-like asteroseismic targets, with
about 100 eventually getting continuous coverage over the entire
mission. This last element is crucial, in that it is only from
extended observations that certain scientifically rich, but extremely
subtle, elements of the oscillation spectra can be measured, e.g.,
frequency asymmetries of mode peaks (diagnostics of granulation) and
frequency asymmetries of mode splittings (diagnostics of magnetic
dynamo signatures).

The first phase of Exercise\,\#1 is concerned with testing extraction
of estimates of the large and small frequency spacings of the
low-degree modes. These seismic frequency spacings will provide key
results for complementing the exoplanet search data (e.g., see Stello
et al. 2007a): the large spacings provide tight constraints for
estimating radii of the exoplanet host stars, which may then be used
to give precise estimates of the exoplanet radii; while the small
spacings may be used to help constrain ages of the host stars. The
ultimate goal of Exercise\,\#1 will be to estimate the radii of the
artificial stars, using the estimated spacings as one of the key data
inputs.

When mode peaks observed in the power frequency spectra are of
sufficient quality it will be possible to obtain extremely precise
estimates of \emph{individual} mode frequencies, for example by
application of peak-bagging fitting methods developed for
Sun-as-a-star data. Use of individual mode frequencies will then allow
even tighter constraints to be placed on the stellar radii, and we
will also test such analyses using the artificial asteroFLAG datasets.

Our intention is to provide input and guidance to the KASC to help
develop a robust strategy, or recipe, for estimation of stellar radii
under different dataset scenarios (i.e., for different intrinsic
stellar properties and noise levels). In this paper we show initial
results from Exercise\,\#1 on estimation of the large spacings. WJC
and four other colleagues (T.~Appourchaux, GH, RN and TS) have thus
far acted as the hares, and have generated artificial data on a
selection of main-sequence stars for other members of the group, the
hounds, to analyse blind. The hounds who have returned results thus
far are T.~Arentoft, JB, OC, STF, RAG, SJJ-R, CR and DS.

 \subsection{The asteroFLAG simulator}
 \label{sec:sim}

\begin{figure*}
 \centerline {\epsfxsize=12.5cm\epsfbox{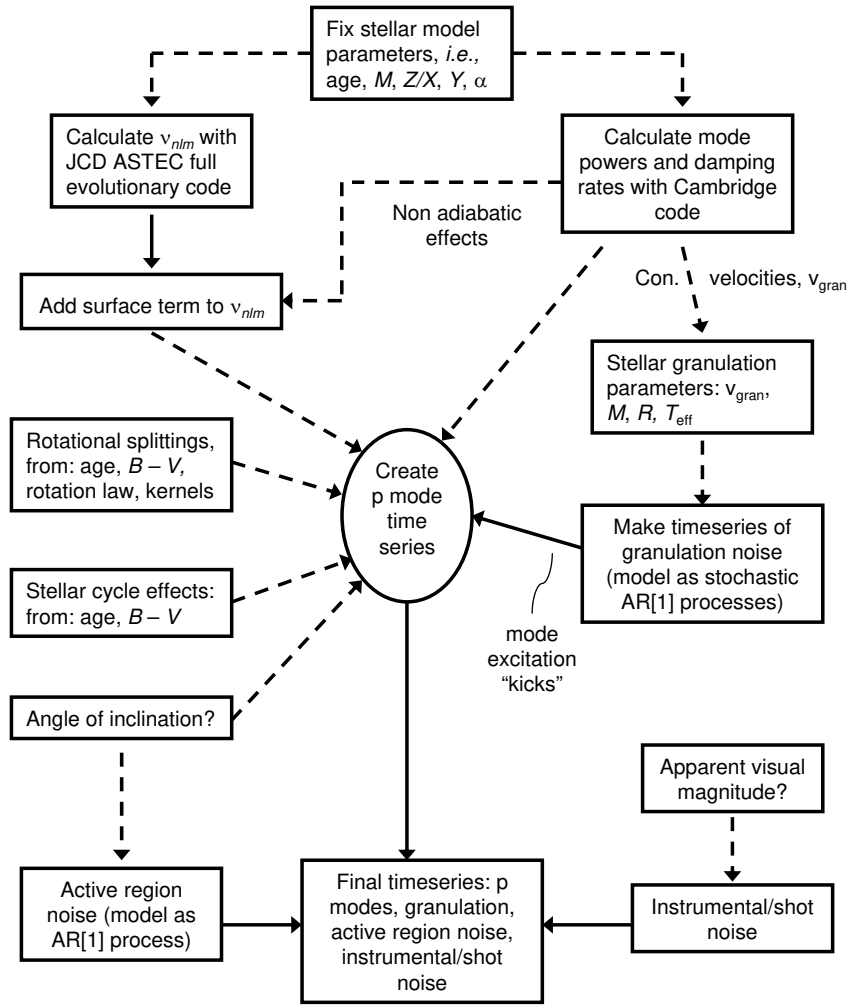}}
 \caption{The main elements of the asteroFLAG simulator.}
 \label{fig:sim}
\end{figure*}

A detailed description of the asteroFLAG simulator will be presented
elsewhere (Chaplin et al., in preparation). Here, we note briefly the
main elements of the simulator. These elements are shown visually in
block diagram form in Figure~\ref{fig:sim}, with the simulator
configured for observations made in intensity. Each timeseries
comprises photometric perturbations from: p modes; granulation noise;
active region noise; and photon shot and instrumental noise.


\subsubsection{Fixing the input parameters}
\label{sec:fixinputs}

We have used the Aarhus stellar evolution code (ASTEC) and the
adiabatic pulsation code ADIPLS to compute theoretical adiabatic
pulsation frequencies of the artificial stars (Christensen-Dalsgaard
2007a, b). We use results on stellar equilibrium and pulsation
calculations, which have been applied in detail to the Sun and other
Sun-like stars (Balmforth 1992; Houdek et al. 1999; Houdek \& Gough
2002), to estimate the excitation rates and damping rates of the p
modes. The amplitudes of the stochastically excited modes were
obtained in the manner of Chaplin et al. (2005, 2007a).

Latitudinally dependent surface rates of rotation were fixed according
to empirical laws, with scaling fixed by the colour, $B-V$, and age of
the star (Aigrain, Favata \& Gilmore 2005; Cardini \& Cassatella 2007;
Donahue et al. 1996; Reiners \& Schmidt 2003). (The $B-V$ colour was
derived from the effective temperature, $T_{\rm eff}$, and metallicity
of the model, using the empirical relations of Alonso, Arribas \&
Mart\'inez-Roger 1996.) Rotation kernels calculated from the stellar
models, together with simple internal rotation laws, were then used to
determine the rotational frequency splittings. We note that the
rotational frequency splittings of the models used in Exercise\,\#1
were fixed deliberately at Sun-like, twice Sun-like and three-times
Sun-like values (see Section~\ref{sec:data} below).

Stellar-cycle effects were not included in the artificial timeseries
data described in this paper. However, the asteroFLAG simulator can
include these effects. The outline implementation is as follows. A
commonly used indicator of surface activity on stars is the Ca~II H \&
K index. This index is usually expressed as $R'_{\rm HK}$: the average
fraction of the star's total luminosity that is emitted in the H \& K
line cores (having been corrected for the photospheric component). We
use scaling relations due to Noyes (1983) and Noyes et al. (1984),
which require $B-V$ and the surface rate of rotation, to estimate
$R'_{\rm HK}$ for our artificial stars. Scaling laws, based on fits to
observations (Radick et al. 1995; Saar \& Brandenburg 2002) of
stellar-cycle variability in $R'_{\rm HK}$ were then used to fix the
amplitudes, $\Delta R'_{\rm HK}$, and the periods of the stellar
cycles. Cycle amplitudes $\Delta R'_{\rm HK}$ were then converted to
equivalent p-mode parameter shifts (e.g., in frequency, power and
damping) in the manner described in Chaplin et al. (2007a). There is
also the provision to include effects of acoustic asphericity, from
the non-homogeneous distribution of the simulated near-surface
activity (again, see Chaplin et al. 2007a).

Mode visibilities were calculated according to Christensen-Dalsgaard
\& Gough (1982). We also allowed for limb darkening, and used the
limb-darkening laws of van~Hamme (1993) to describe variation of the
limb-darkening properties from one star to another. These laws use the
effective temperature, $T_{\rm eff}$, and the surface gravity to fix
the coefficients describing the limb-darkening.

To estimate the standard deviation and timescale of the granulation,
we have used scaling laws, and scaled against solar values (see also
Stello et al. 2007b; Kjeldsen \& Bedding, in preparation). The solar
values were estimated from analysis of a multi-year timeseries of
solar intensity observations made by the VIRGO/PMO6 instrument on
board the \emph{ESA/NASA} SOHO spacecraft. We have assumed that the
standard deviation of the brightness fluctuations scales in inverse
proportion to the square root of the number of convective cells on the
surface of the star (we assume the cells are statistically
independent; see also Ludwig 2006). The characteristic size of cells
was assumed to scale with the product of the mixing length parameter
and the isothermal scale height at the surface of the star. The
brightness fluctuations associated with individual cells were assumed
to scale with the surface velocities associated with the convection,
which came from the stochastic excitation models that were used to
predict the mode excitation and damping rates. To estimate the
timescale of the granulation we assumed it scales in inverse
proportion to the acoustic cut-off frequency of the star.

The standard deviation of the active region noise was assumed to scale
with $R'_{\rm HK}$, in the manner described by Aigrain, Favata \&
Gilmore (2005).  We also allowed for the effect of the angle of
inclination of the star, having assumed the active-region signal was
confined within certain bands of stellar latitude, using the work of
Knaack et al. (2001) as a guide. We assumed the timescale of the
active-region signal scaled with the period of rotation of the star.

For the photon shot noise, we used the latest estimates of the
expected precision of the Kepler observations at different
$m_v$. These data were provided to HK by the Kepler team.

\subsubsection{Generation of the artificial data}
\label{sec:gen}

Components due to granulation and active regions were generated in the
time domain using a stochastic, autoregressive model of the AR[1] type
(e.g., see Koen 2005).

We used code based on the latest version of the solarFLAG (Chaplin et
al. 2006) timeseries generation code to make the p-mode timeseries
(the solarFLAG generator is described in Chaplin et al., in
preparation). Each ($n,l,m$) component was generated in the time
domain. Components were modeled as forced, damped harmonic
oscillators, re-excited at frequent intervals (here, every
minute). The oscillators were re-excited by a timeseries of `kicks'
made from the granulation noise. Kicks given to overtones of the same
($l,m$) were correlated. Implicit in this approach is the assumption
that on the Sun, the excitation function of a mode of a given ($l,m$)
is the same as the component of the granulation that has the same
spherical harmonic projection over the corresponding range in
frequency (Toutain, Elsworth \& Chaplin 2005).  Correlations of the
excitation of different overtones, coupled with the later addition of
granulation noise to the p-mode timeseries, gave rise to asymmetries
of the resonant peaks in the power frequency spectrum (asymmetries
that are displayed by the solar p modes).

\begin{figure}[h]
 \centerline
 {\epsfxsize=9.0cm\epsfbox{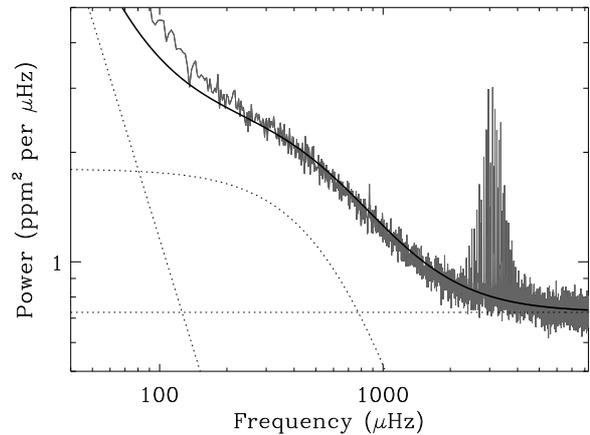}}

 \caption{How Kepler would see the Sun, according to the asteroFLAG
  simulator, were it to be observed continuously for 4\,yr at apparent
  visual magnitude $m_v=9$. The power frequency spectrum of the
  timeseries is rendered in light grey, with the most prominent modes
  observed at frequencies around $\sim 3000\,\rm \mu Hz$.  The
  individual limit contributions of photon shot noise, granulation,
  and active-region noise are shown as the dotted lines. The total
  contribution of these three ``noise'' components is plotted as a
  dark, continuous line.}

 \label{fig:sun09}
\end{figure}

To give an idea of the expected quality of the Kepler observations,
Figure~\ref{fig:sun09} shows how Kepler would see the Sun, according
to the asteroFLAG simulator, were the Sun to be observed continuously
for 4\,yr at apparent visual magnitude $m_v=9$. This is at the bright
$m_v$ end of the target range, which extends down to brightness $m_v
\sim 15$ (e.g., Christensen-Dalsgaard et al. 2007).


 \subsection{Data for Exercise\,\#1}
 \label{sec:data}

For the first phase of Exercise\,\#1, we have prepared timeseries of
Kepler-like data on 3 artificial stars. We have named the artificial
stars Pancho, Boris and Katrina\footnote{The chosen names are names of
cats with whom several of the authors are acquainted. Given the
origins of the naming of our stars, it has been suggested that our
hare-and-hounds exercises should instead be called cat-and-mouse
exercises.} Here, we present results on a total of twenty-eight 4-yr
timeseries prepared for each star. These timeseries covered all
possible combinations of the following:
\begin{itemize}

 \item Four different apparent visual magnitudes, $m_v=9$, 11, 13 and
 15;

 \item Three different angles of stellar inclination, $i=0$, 30 and 60
 degrees;

 \item Three different mean internal rates of rotation: Sun-like,
 twice Sun-like, and three-times Sun-like;

\end{itemize}
We note that the same realization noise was always used to excite the
overtones of each ($l,m$) of a given star, regardless of the dataset;
however, different realizations of the photon shot noise were added at
each of the different apparent magnitudes.  As indicated previously,
the timeseries did not include stellar-cycle-like effects.

These multiple timeseries were intended to help the hounds test their
analysis codes, and so the hares also passed to the hounds a priori
information on the inclinations and rotation rates of the datasets.
It is in the next phase of Exercise\,\#1 that data will be released at
only one $m_v$ per star, on a selection of new stars, thereby
providing true blind tests for the hounds. We comment below on the
potential impact on the results of the additional a priori information
given by having multiple timeseries on each star.


\begin{figure}[h]
 \centerline
 {\epsfxsize=8.3cm\epsfbox{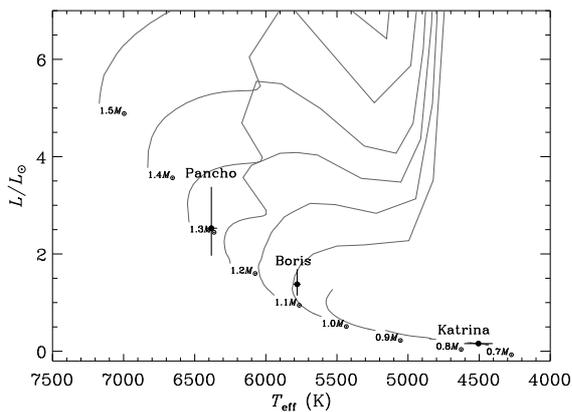}}

 \caption{Locations of the stars Pancho, Boris and Katrina (see figure
 annotation) on a temperature-luminosity diagram, with locations
 inferred from the traditional data released on the stars. Here, we
 used the artificial parallax data for $m_v=9$. The grey lines show
 evolutionary tracks, made with the Padova models (Bonatto, Bica \&
 Girardi 2004; Girardi et al. 2002, 2004).}

 \label{fig:where}
\end{figure}


In addition to the timeseries data, the hares also released
`traditional' data on the stars, appropriate to each apparent visual
magnitude, $m_v$. The traditional data were: parallaxes, $\pi$,
metallicities, [Fe/H], and effective temperatures, $T_{\rm eff}$.
(Full details on these data will be given in a separate paper.)
Figure~\ref{fig:where} shows the location of the artificial stars on a
temperature-luminosity diagram. The positions have been inferred using
the traditional data (with the parallax for $m_v=9$). Here, we
followed procedures the hounds would need to adopt to convert from
absolute visual to absolute bolometric magnitude (we used the
bolometric corrections of Flower 1996). The locations of the stars in
Figure~\ref{fig:where} indicate that Pancho is intrinsically the
brightest star; whilst Katrina is intrinsically the faintest star.

\begin{figure*}[h]
 \centerline {\epsfxsize=6.0cm\epsfbox{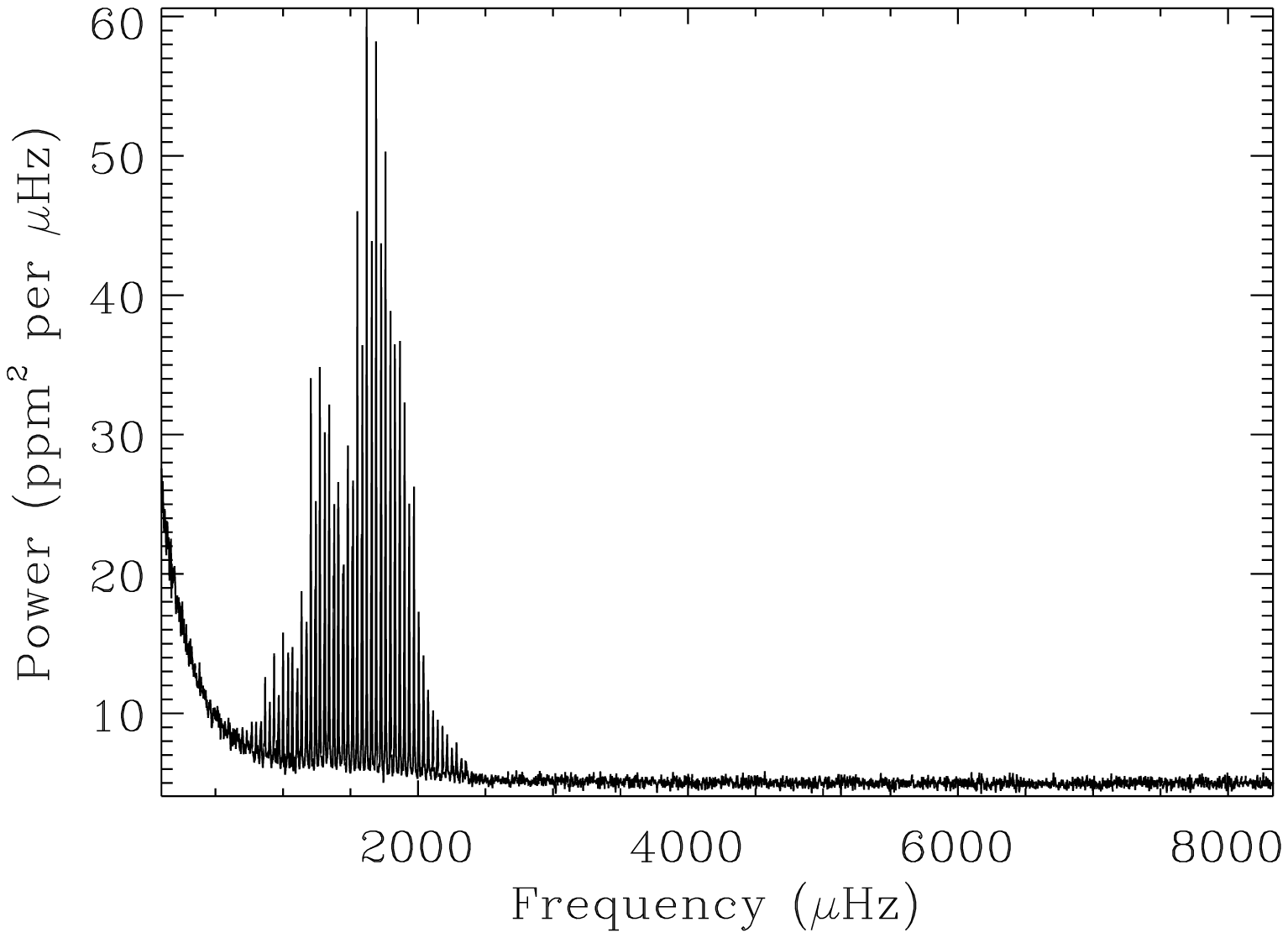}
 \epsfxsize=6.0cm\epsfbox{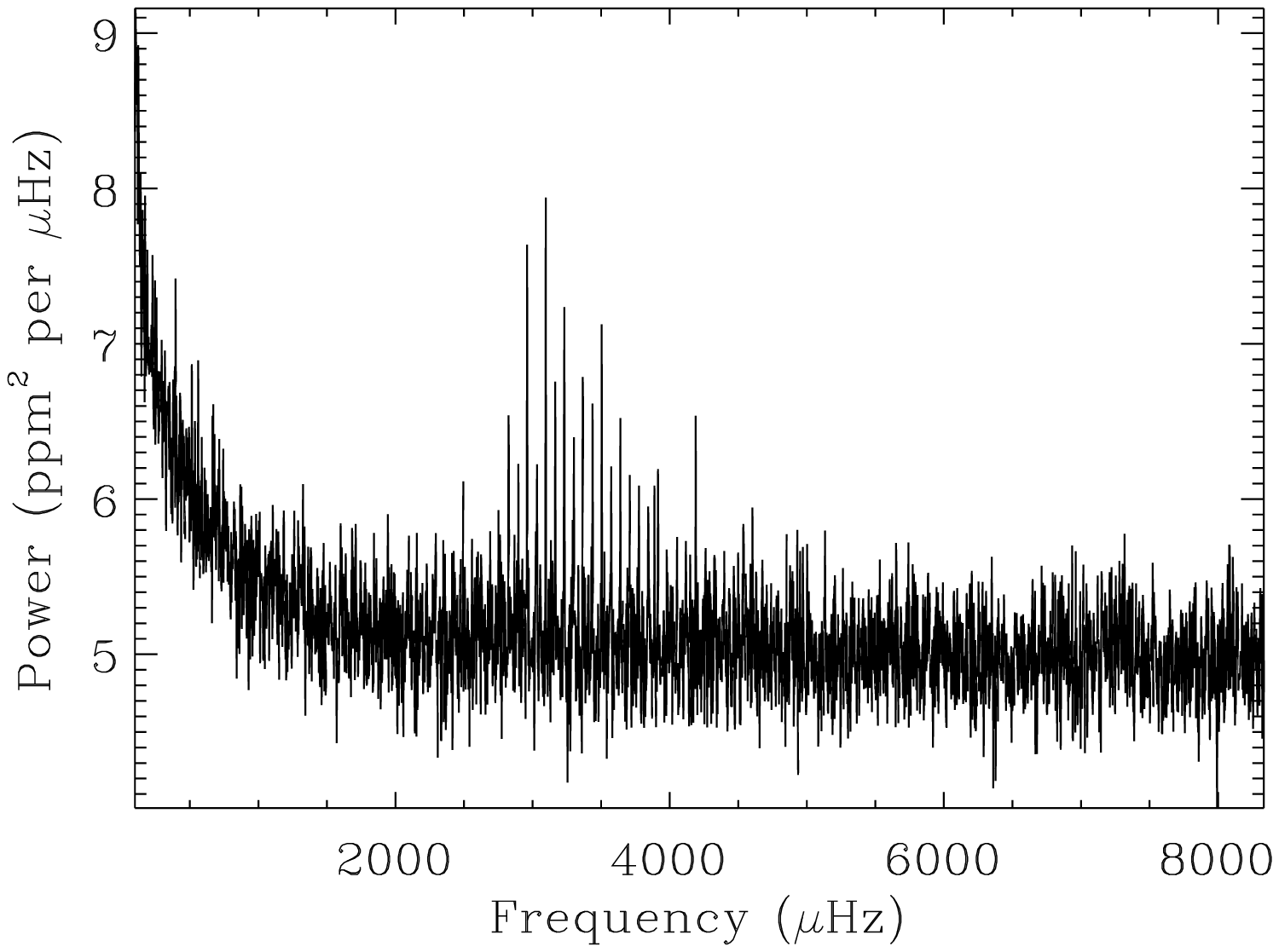}
 \epsfxsize=6.0cm\epsfbox{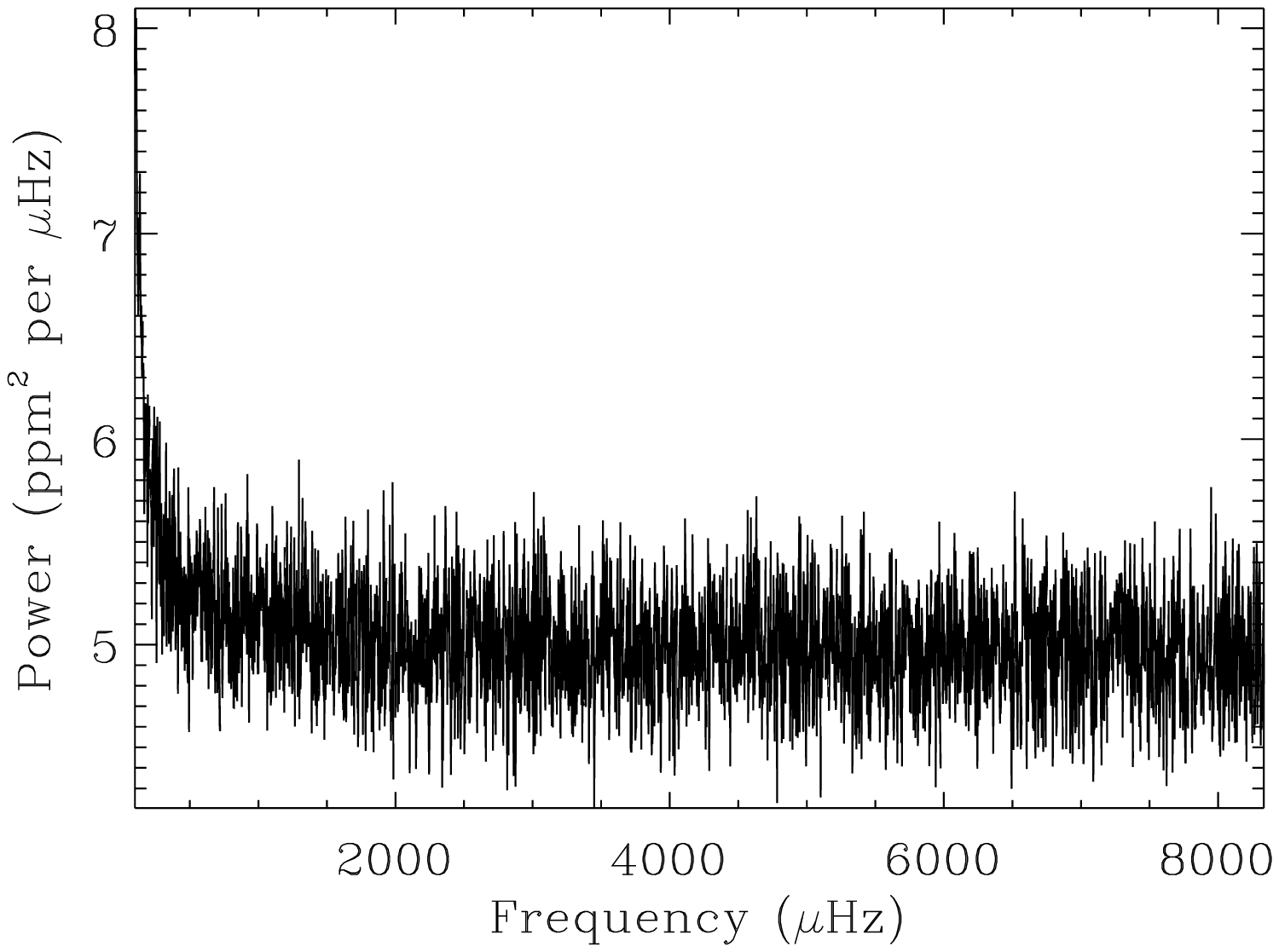}}

 \caption{How Kepler would see Pancho (left-hand panel), Boris (middle
  panel) and Katrina (right-hand panel), according to the asteroFLAG
  simulator, were the artificial stars to be observed continuously for
  4\,yr at apparent visual magnitude $m_v=11$.  The plots show average
  power frequency spectra, made by dividing the full timeseries into
  short 4-day segments and co-adding the individual spectra.}

 \label{fig:starpow}
\end{figure*}

Figure~\ref{fig:starpow} shows how Kepler would see Pancho (left-hand
panel), Boris (middle panel) and Katrina (right-hand panel), according
to the asteroFLAG simulator, were the artificial stars to be observed
continuously for 4\,yr at apparent visual magnitude $m_v=11$. Pancho
shows the strongest p-mode amplitudes of the three stars, while
Katrina has the weakest modes.


\section{Results}

The hounds made up six teams: JB, RAG and SJJ-R formed one team; while
T.~Arentoft, OC, STF, CR and DS worked individually. The hounds
extracted estimates of the large frequency spacings using various
methods. T.~Arentoft used a Matched Filter approach, which is similar
to methods applied in searches for exoplanet transits (see
Christensen-Dalsgaard et al. 2007). The other hounds used either
Fourier transforms of the power frequency spectra (one hound actually
worked on autocorrelation functions of the timeseries) or
autocorrelations of the power frequency spectra (e.g., see Scargle
1989; Ransom, Eikenberry \& Middleditch 2002).  We also note that JB,
RAG and SJJ-R worked on ``denoised'' power frequency spectra of
timeseries, using the curvelet transform (Lambert et al. 2006).  For
analysis of several timeseries, some hounds divided the timeseries
into shorter pieces. Power frequency spectra of these shorter pieces
were then computed and co-added for subsequent analysis.

In all cases the hounds had to make choices over what ranges in
frequency to use for the analyses. Estimates of the large spacings are
affected to some extent by the range that is chosen, because the large
spacings are frequency dependent. The angle of inclination and the
magnitudes of the mode frequency splittings -- both of which affect
the appearance of the mode peaks, both within multiplets, and from one
degree $l$ to another -- will also have a subtle effect on the
results.  The estimates may also be affected by the implementation of
the analysis, for example from the manner in which peaks in the
autocorrelation, or Fourier transform, of the power frequency spectrum
are fitted to yield the estimated spacings. Before we show the main
results of the hounds, we first illustrate the impact of some of the
effects listed here.

We computed the smooth, noise-free ``limit'' power frequency spectrum
of each timeseries, and then analysed each noise-free power frequency
spectrum to estimate the large spacings. When we analysed each
spectrum, we used the range in frequency that each hound chose when
they analysed that same spectrum.  We estimated the spacings using
both of the basic methods adopted by the hounds. The results that are
given correspond to a ``best case scenario'', in that there is no
noise in the power frequency spectrum. We note that our implementation
of the methods most likely differed in detail from implementations of
the hounds. The results provide a guide to differences we might expect
from the free choice of the frequency range and analysis method, and
of the inclination and mode frequency splittings.  It is worth adding
that the frequency range selected for analysis is information one
would aim to use in determining the stellar radius, even if it was not
possible to measure individual mode frequencies reliably.
 
\begin{figure*}
 \centerline {\epsfxsize=7.0cm\epsfbox{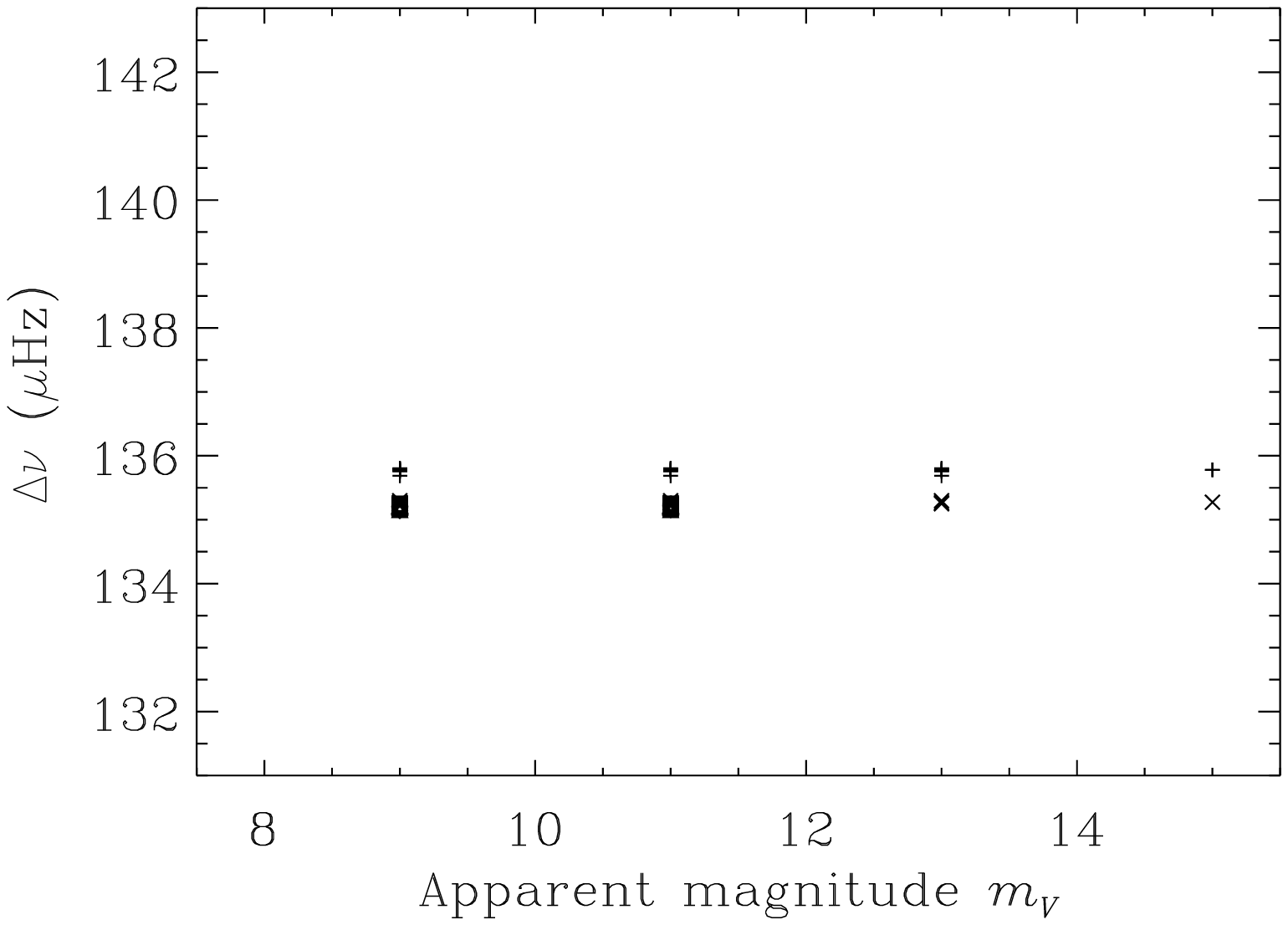}
 \epsfxsize=7.0cm\epsfbox{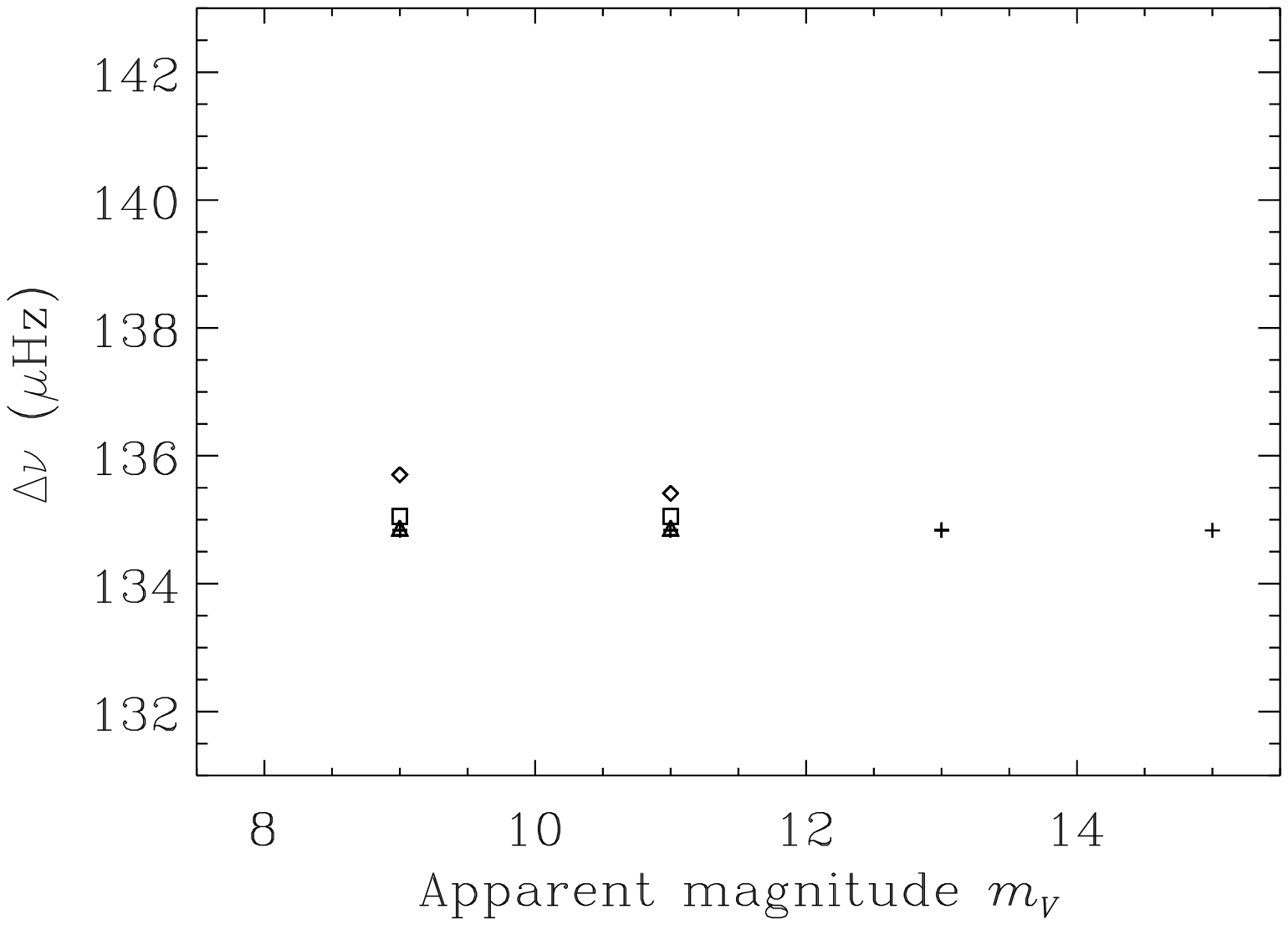}}

 \caption{Estimated spacings extracted from noise-free power frequency
  spectra of the artificial star Boris.  Each symbol is used to show
  results for the frequency ranges chosen by a particular hound. The
  left-hand panel shows results from using the autocorrelation of the
  power frequency spectrum; while the right-hand panel shows results
  from using the Fourier transform of the power frequency spectrum.}

 \label{fig:borislim}
\end{figure*}

Figure~\ref{fig:borislim} shows the estimated spacings extracted from
noise-free power frequency spectra of the artificial star Boris.  Each
symbol is used to show results for the frequency ranges chosen by a
particular hound. The left-hand panel shows results from using the
autocorrelation of the power frequency spectrum; while the right-hand
panel shows results from using the Fourier transform of the power
frequency spectrum. Agreement between the results is of course very
good, so much so that it is hard to tell results apart for the
autocorrelation method. However, there are differences present, which
are in some cases as large as $\approx 0.75$\,\% of the mean
spacings. We consider below how scatter in the large spacing estimates
translates to uncertainty in estimates of stellar radii.


\begin{figure*}
 \centerline
 {\epsfxsize=7.0cm\epsfbox{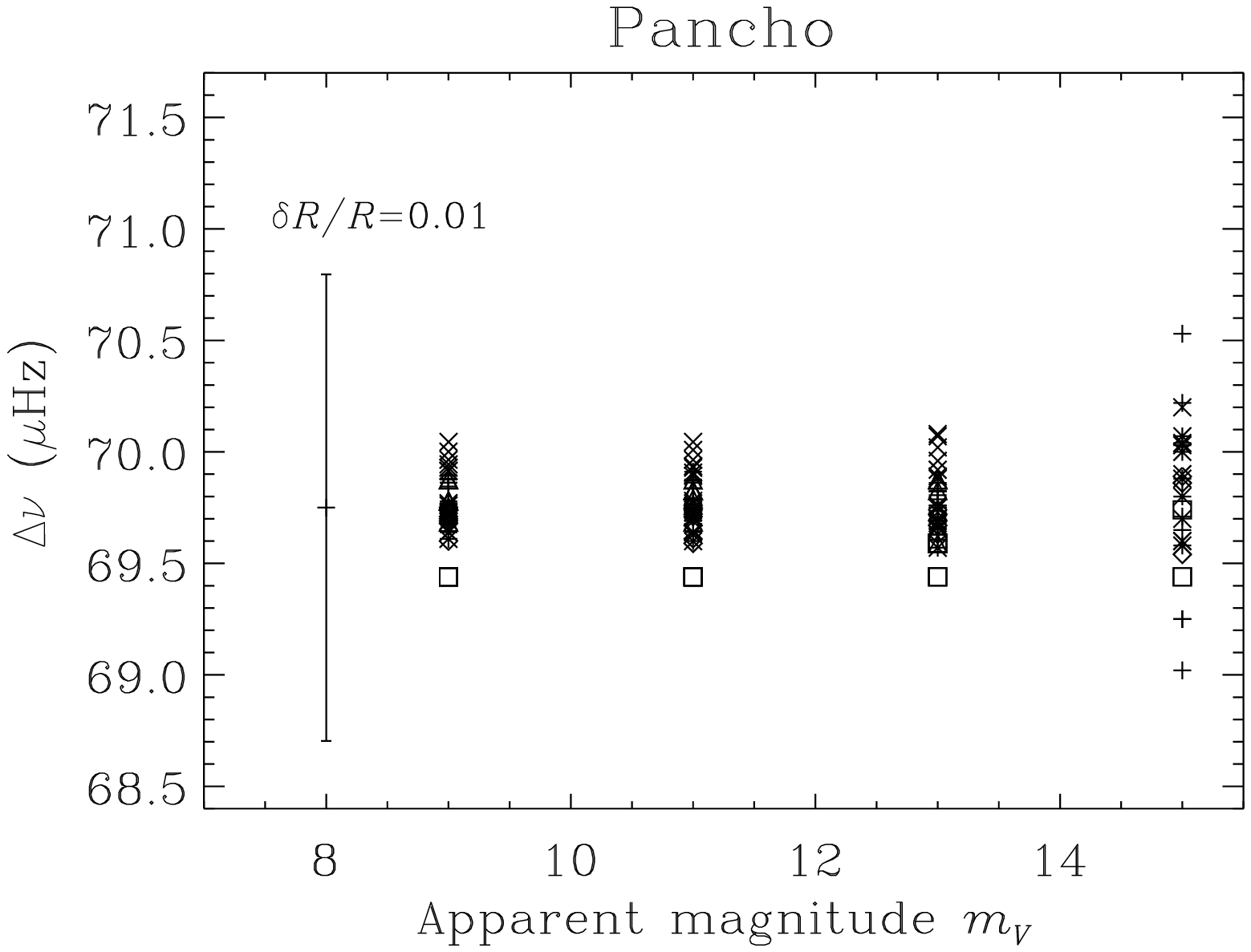}
  \epsfxsize=7.0cm\epsfbox{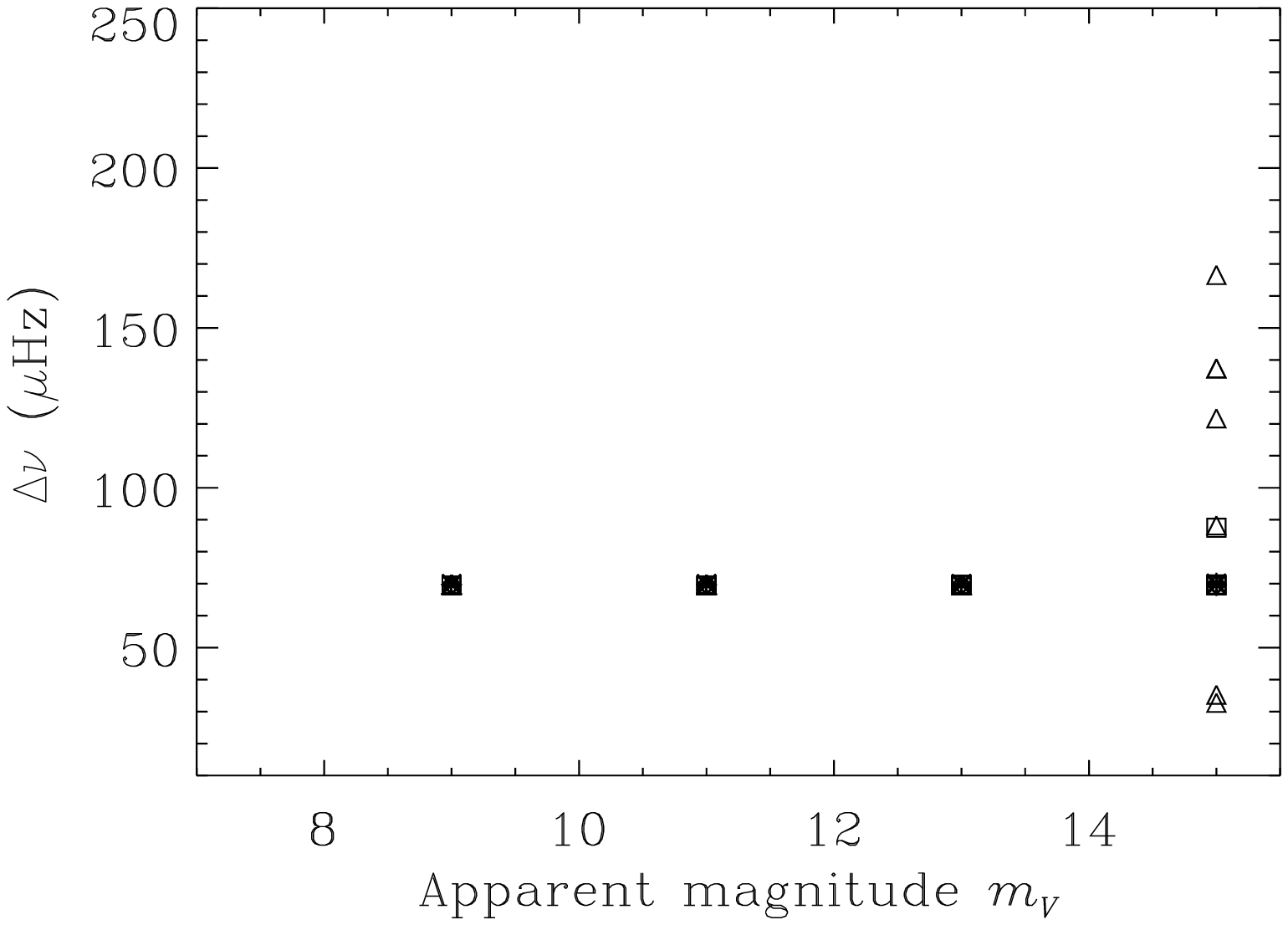}}
 \centerline
 {\epsfxsize=7.0cm\epsfbox{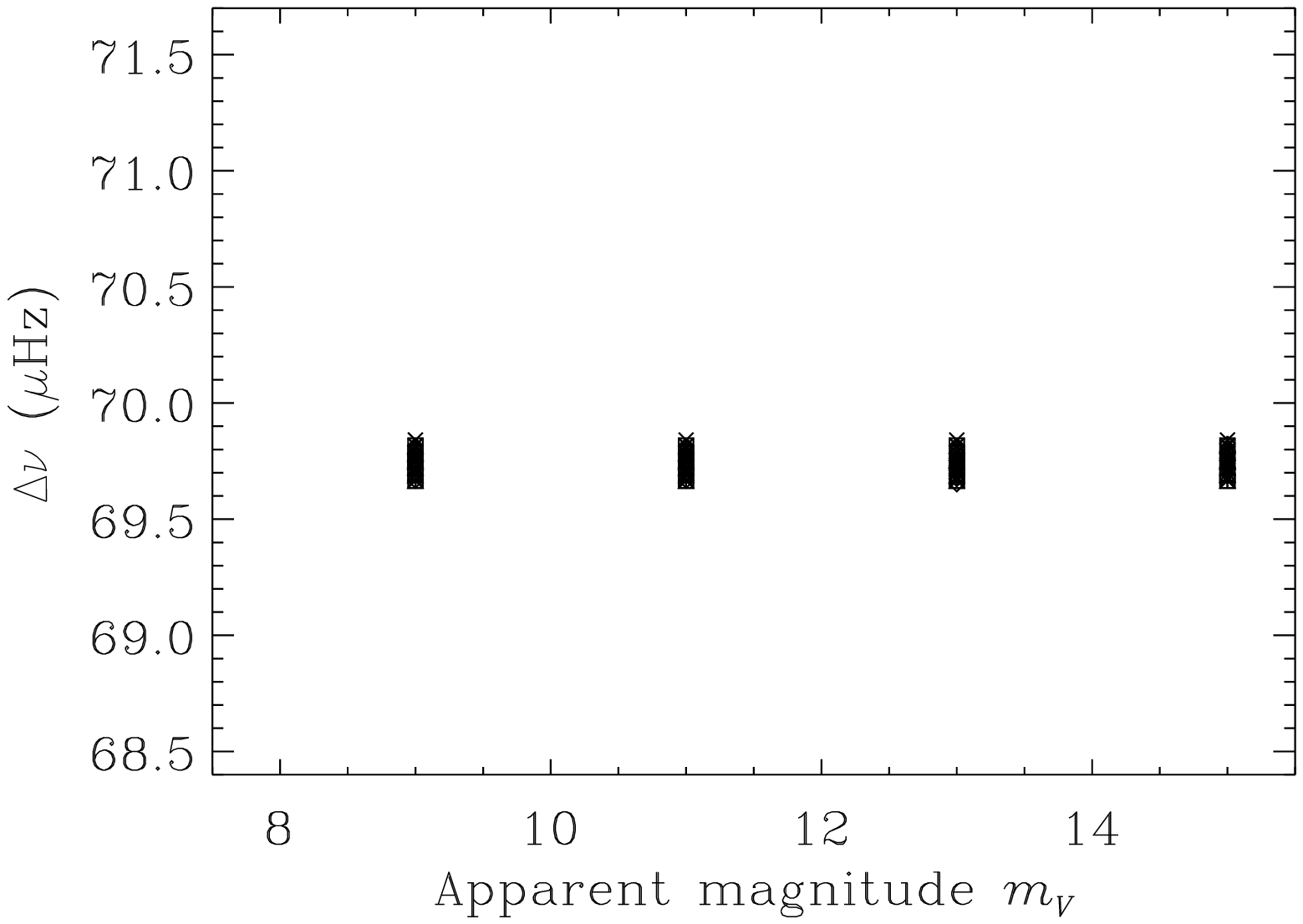}
  \epsfxsize=7.0cm\epsfbox{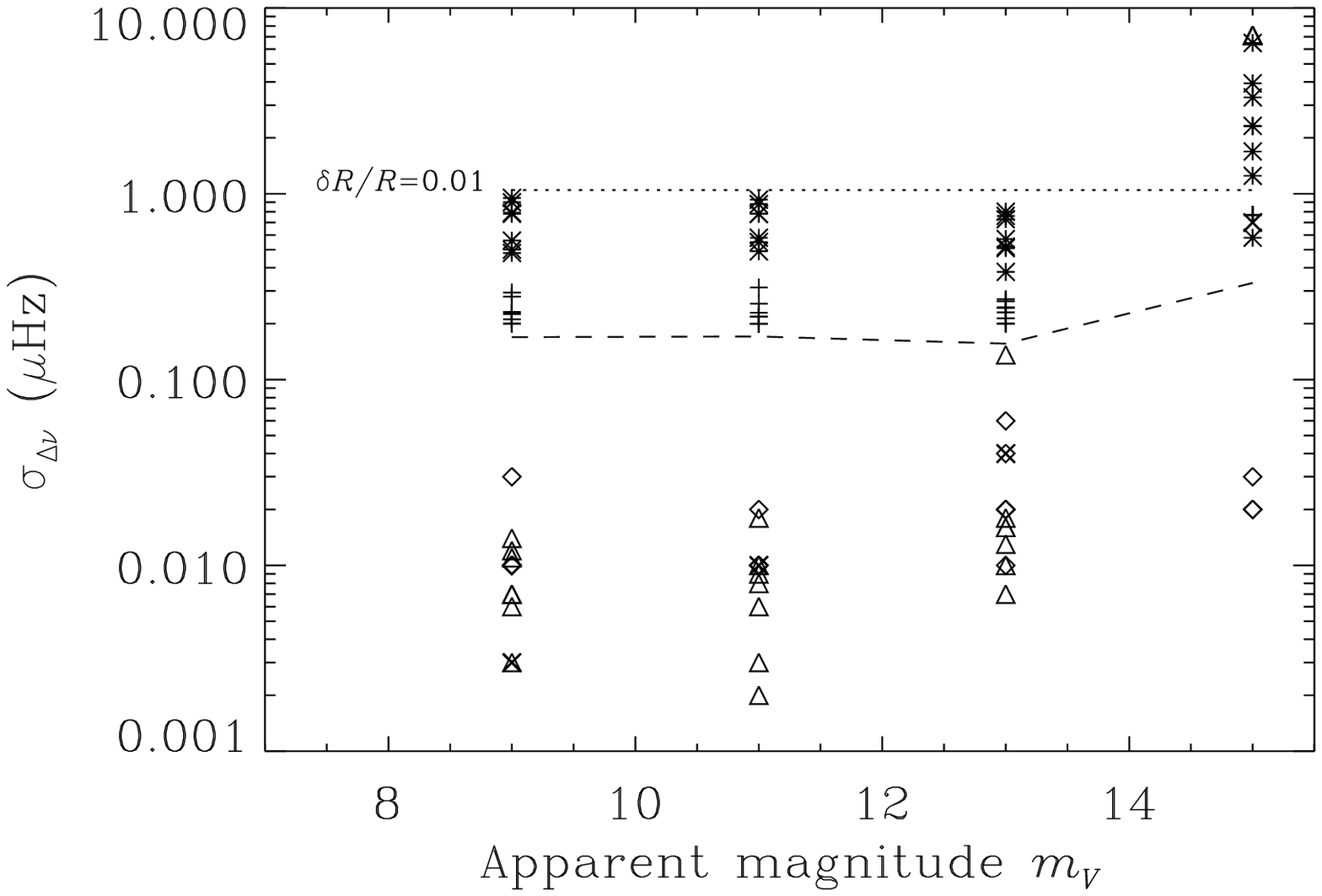}}

 \caption{Results, on estimation of the large spacings, $\Delta\nu$,
 for artificial timeseries of Pancho. Symbols show results for
 different hounds. Top left-hand panel: estimates of the large
 spacings returned by the hounds. The error bar shows the uncertainty
 in $\Delta\nu$ required to give a 1\,\% uncertainty on the inferred
 radius of the star. Top right-hand panel: estimates of the large
 spacings, but plotted over a much larger range on the ordinate to
 show poor estimates of the spacings. Bottom left-hand panel:
 estimated spacings extracted from noise-free power frequency spectra
 by WJC (see Figure~\ref{fig:borislim}), using the autocorrelation of
 the power frequency spectrum.  This is the only panel in which
 symbols do not indicate results for different hounds, but rather the
 results of WJC for the \emph{frequency ranges} chosen by different
 hounds. Bottom right-hand panel: estimates of the uncertainties on
 the large spacings returned by the hounds. The dashed line shows the
 standard deviation of the results of the hounds at each apparent
 visual magnitude, $m_v$. The dotted line again shows the uncertainty
 in $\Delta\nu$ required to give a 1\,\% uncertainty on the inferred
 radius of the star.}

 \label{fig:panchores}
\end{figure*}

\begin{figure*}
 \centerline
 {\epsfxsize=7.0cm\epsfbox{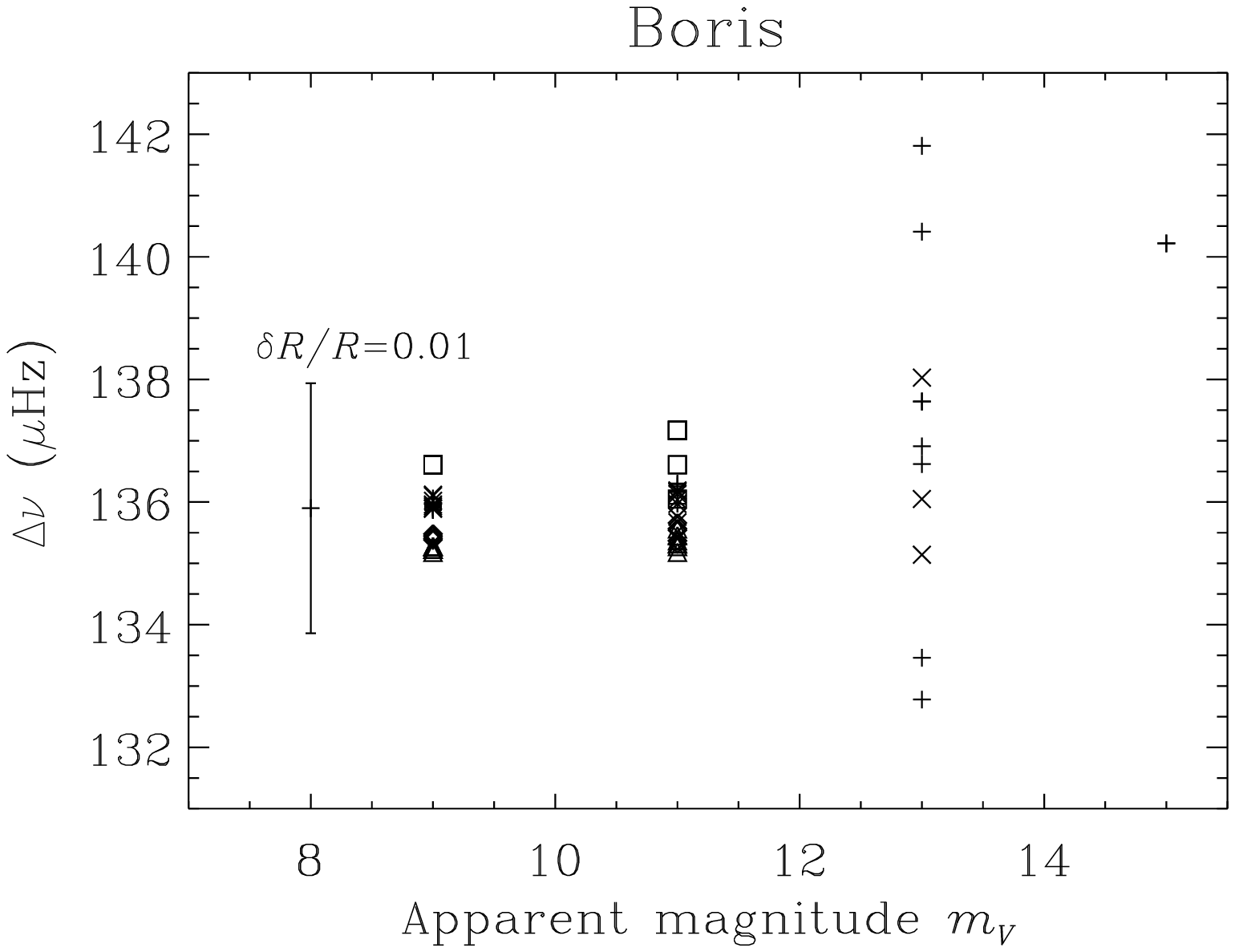}
  \epsfxsize=7.0cm\epsfbox{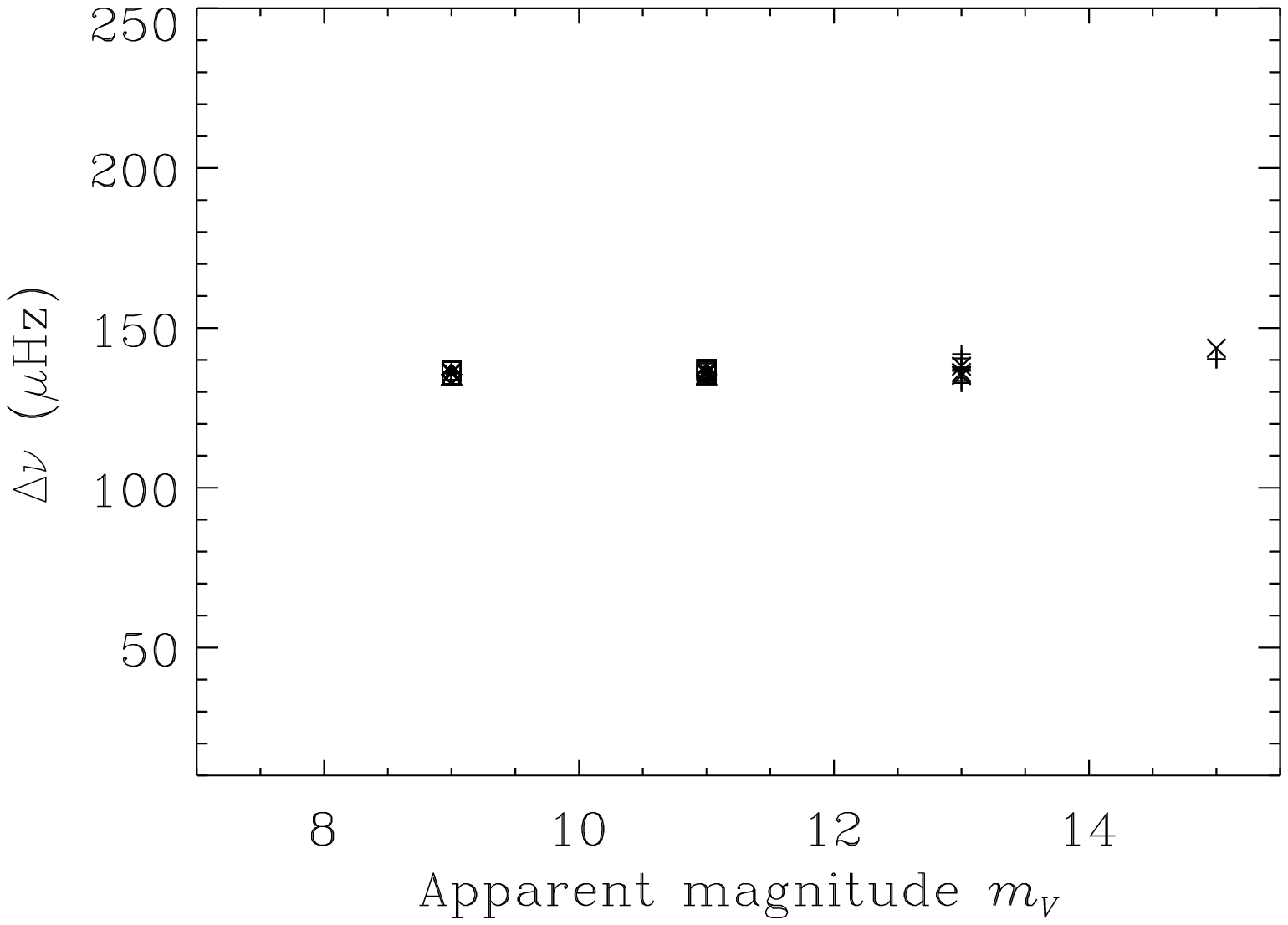}}
 \centerline
 {\epsfxsize=7.0cm\epsfbox{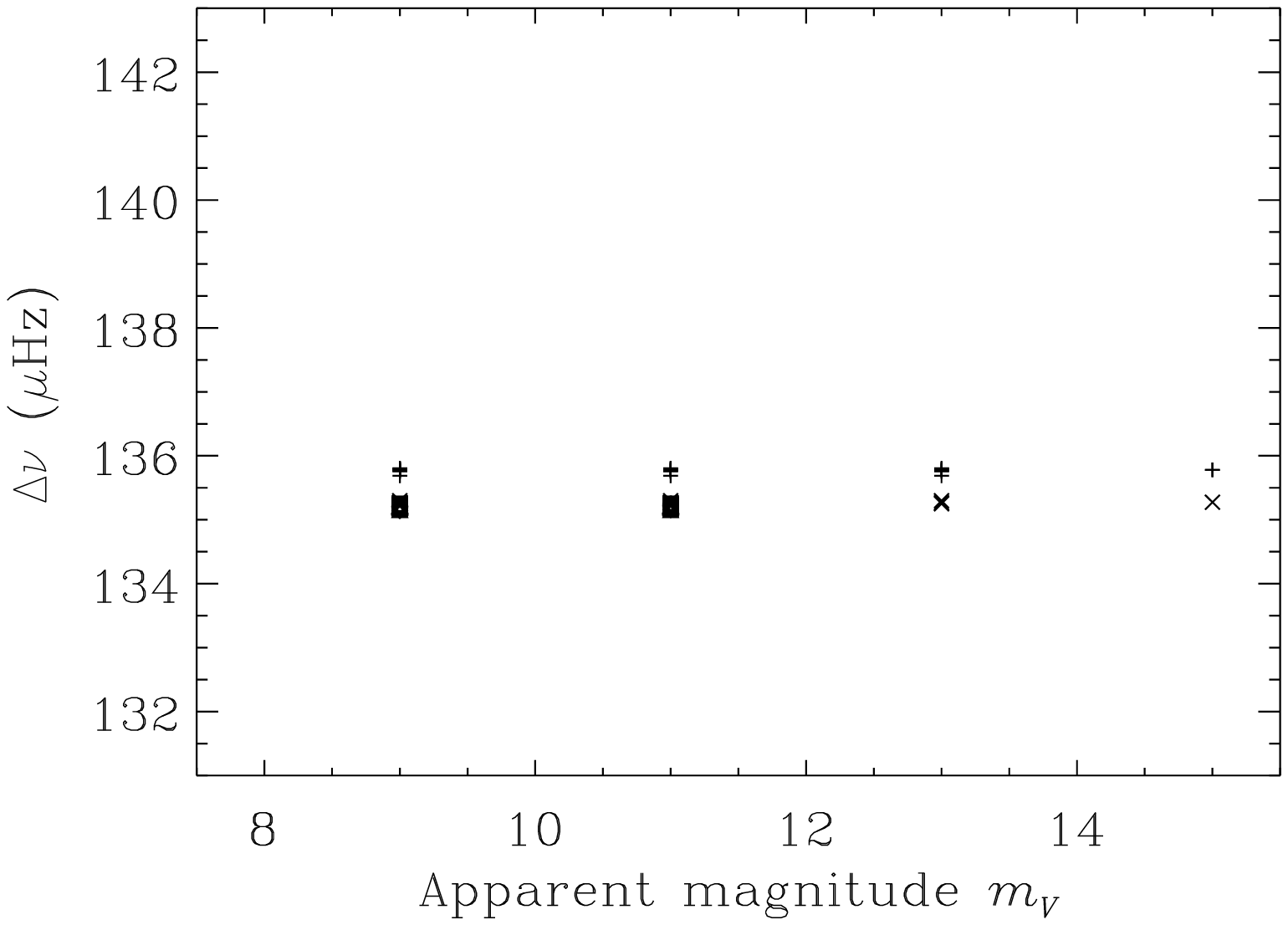}
  \epsfxsize=7.0cm\epsfbox{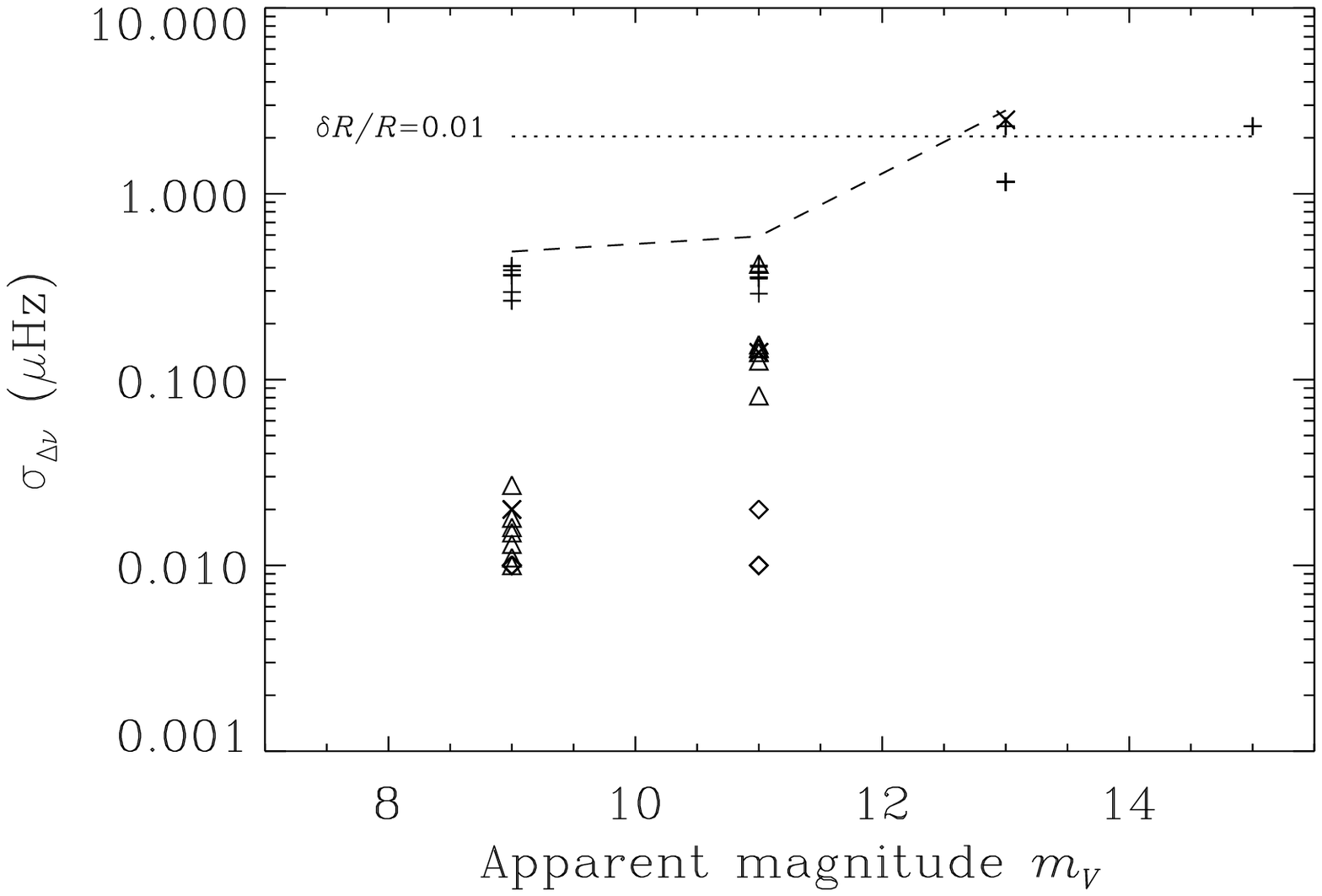}}

 \caption{Results, on estimation of the large spacings, for artificial
 timeseries of Boris. (Details as per Figure~\ref{fig:panchores}.)}

 \label{fig:borisres}
\end{figure*}

\begin{figure*}
 \centerline
 {\epsfxsize=7.0cm\epsfbox{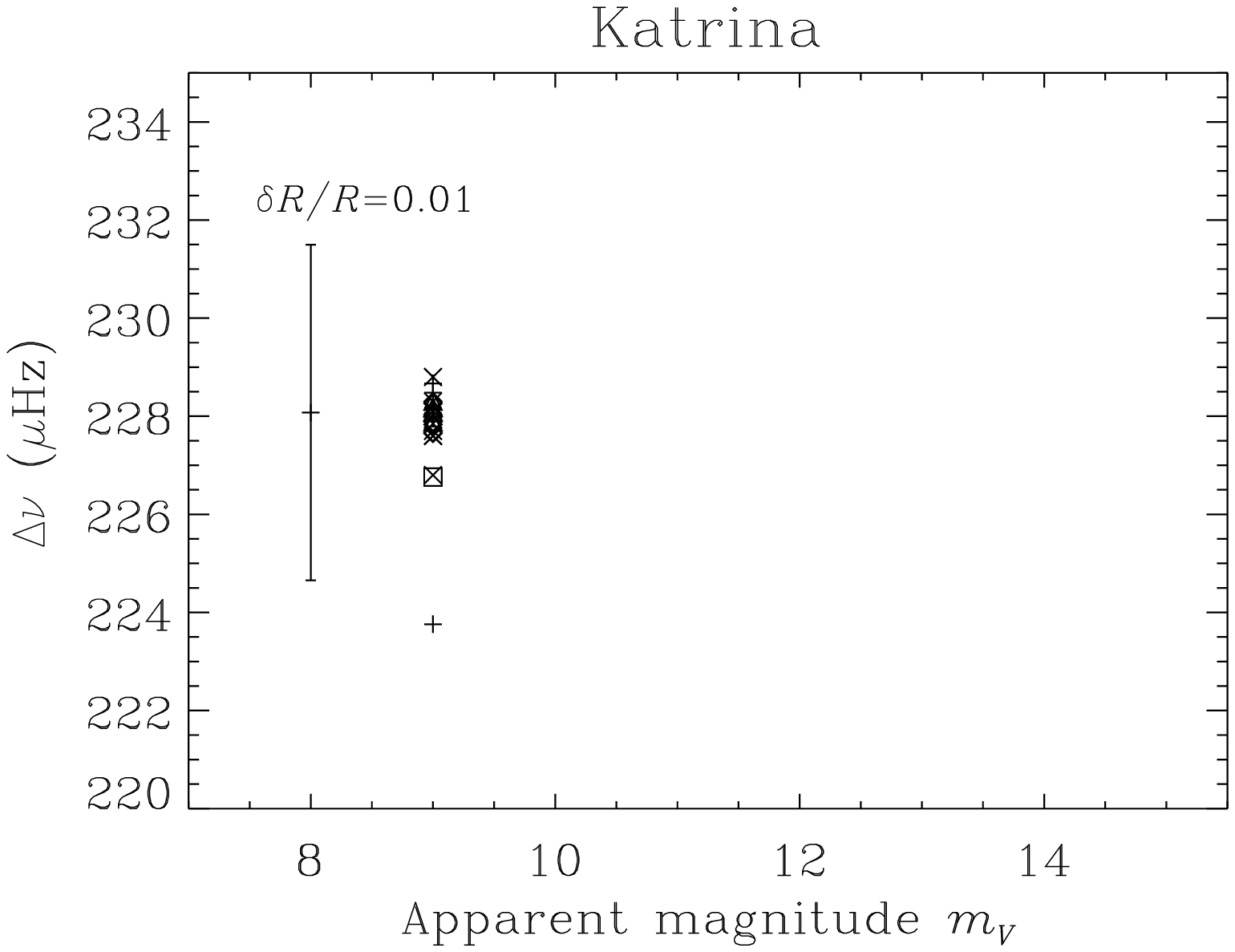}
  \epsfxsize=7.0cm\epsfbox{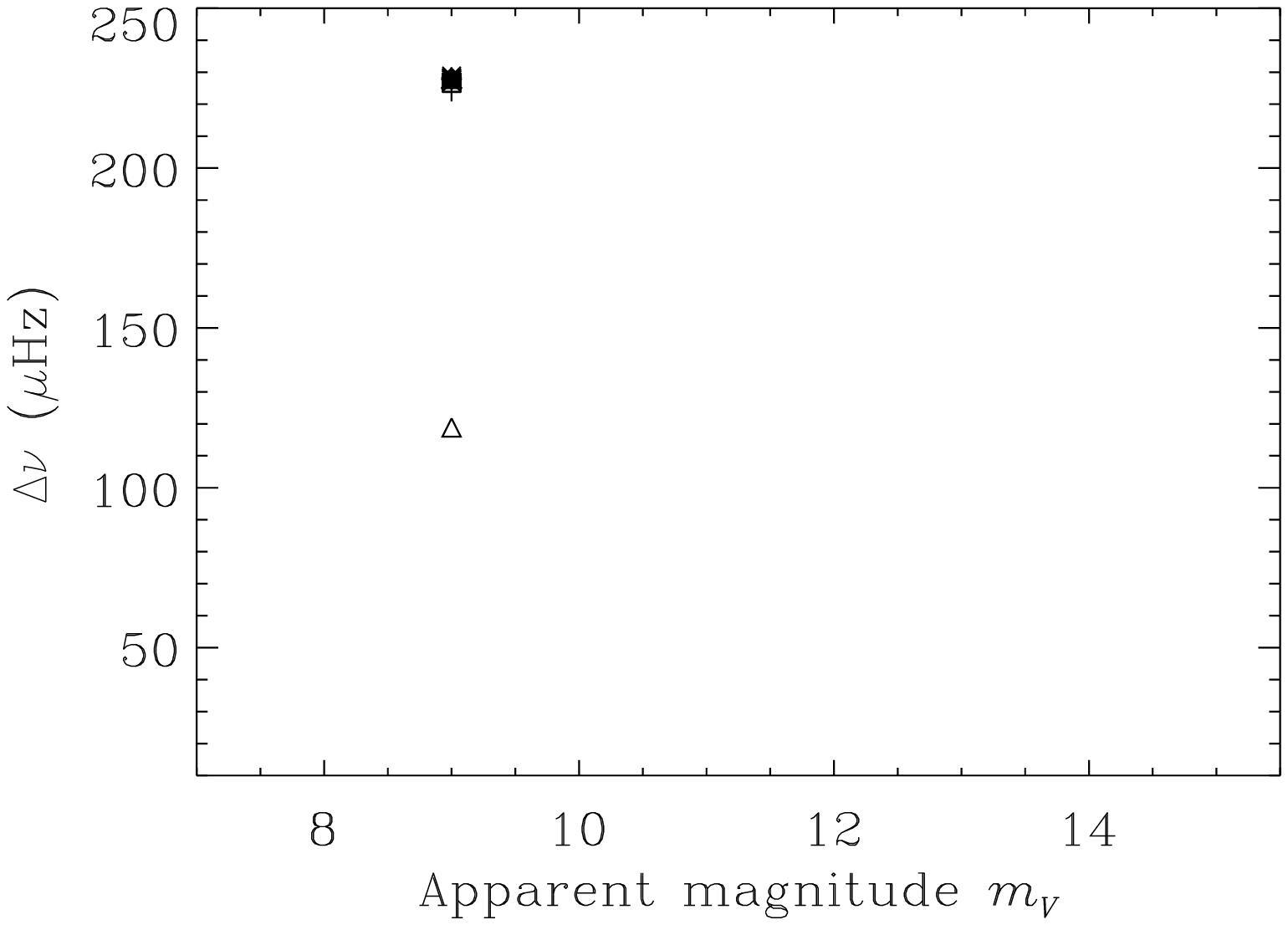}}
 \centerline
 {\epsfxsize=7.0cm\epsfbox{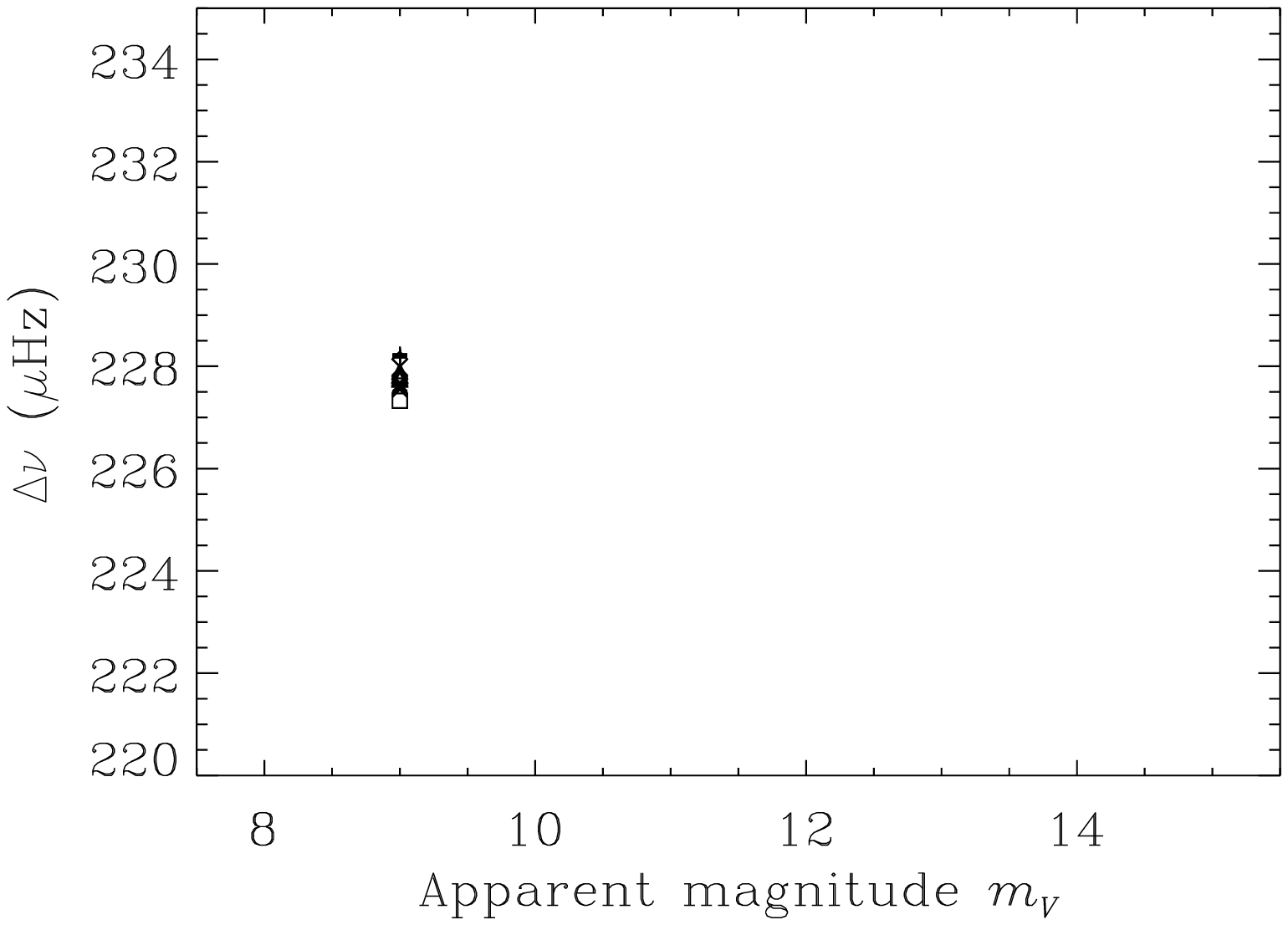}
  \epsfxsize=7.0cm\epsfbox{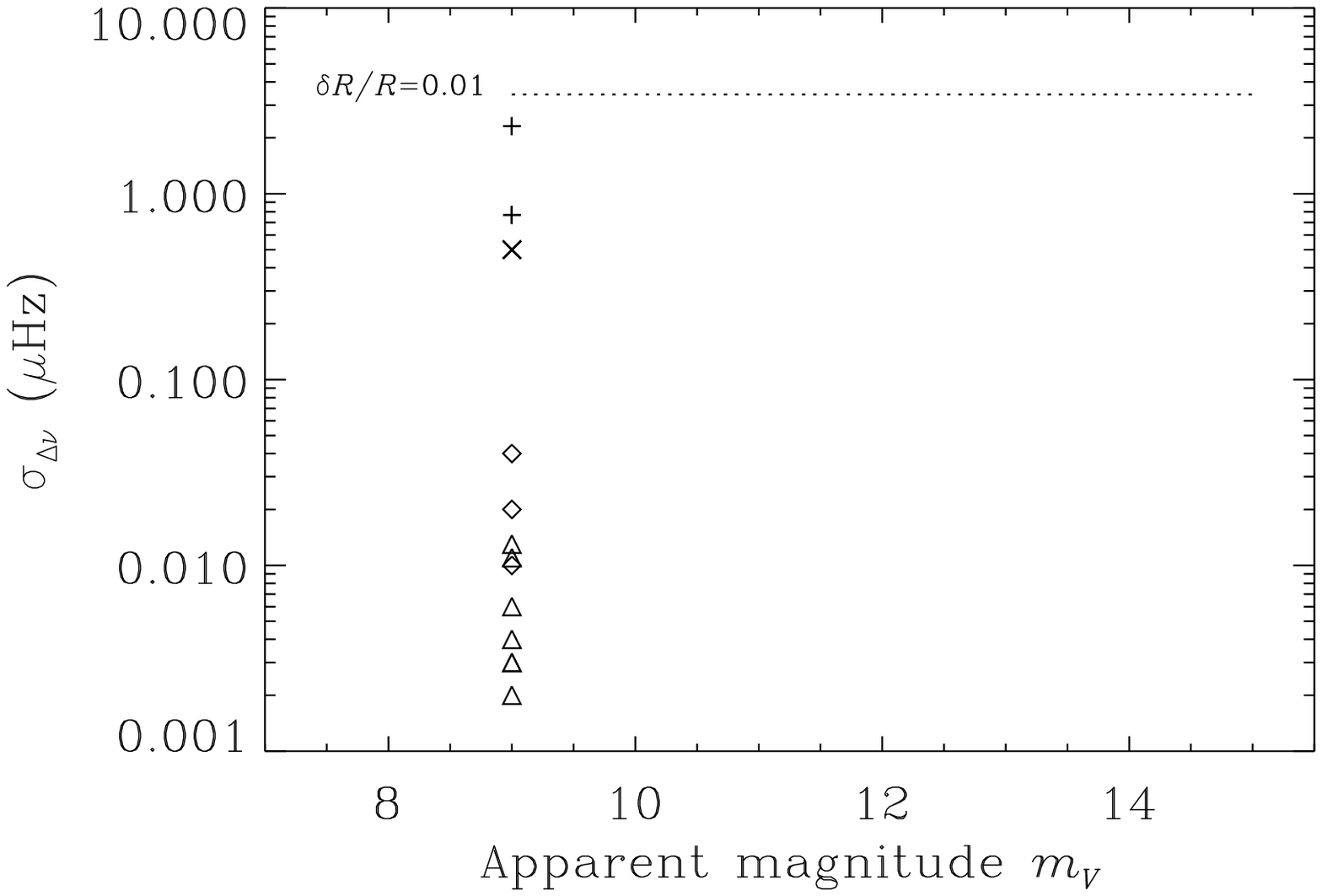}}

 \caption{Results, on estimation of the large spacings, for artificial
 timeseries of Katrina. (Details as per Figure~\ref{fig:panchores}.)}

 \label{fig:katrinares}
\end{figure*}

The main results on the large frequency spacings $\Delta\nu$ are shown
for each star in Figures~\ref{fig:panchores},~\ref{fig:borisres}
and~\ref{fig:katrinares}.  The various symbols show results for
different hounds. (JB, RAG and SJJ-R concentrated on developing
application of the curvelet technique, and have returned results
initially for Pancho only.) A first look at the results showed that
there was no clear, systematic dependence of the estimated spacings on
either the inclination or the internal rate of rotation. We therefore
decided not to differentiate between these parameters when plotting
the results (so that all estimates of $\Delta\nu$ for a given star are
plotted on the same graph).

It is important to recognise that choices made for the higher noise
(larger $m_v$) spectra may have been influenced by results on the
lower noise (smaller $m_v$) spectra. Perhaps the most important choice
is the part of the power frequency spectrum that is selected for
subsequent analysis. The choice can be less than straightforward when
the mode peaks are not readily apparent in the power frequency
spectrum. Under such circumstances it may still be possible to extract
a robust estimate of the spacing provided the right part of the
spectrum is selected. This is the most obvious example of where our
hounds will have learned from, and therefore benefited by, their
selections made for the lower noise spectra.

We should add that information included with the results returns
(together with the results themselves) suggests that the hounds did
attempt to treat each timeseries on its own merits. But this may not
always have been possible where the choice of frequency range was
concerned. For analysis on real stars, and in the absence of prominent
mode peaks in the power frequency spectrum, one would seek to use
already-known traditional information on the star to help fix suitable
ranges in frequency to analyse. Furthermore, the curvelets used by JB,
RAG and SJJ-R, certainly help in cleaning up the raw spectra to reveal
the presence of any modes.


Let us now turn to the main results. The top left-hand panel of each
figure shows estimates of the large spacings returned by the
hounds. The error bars show the uncertainties in $\Delta\nu$ required
to give 1\,\% uncertainties on the inferred radii of the stars. These
values follow straightforwardly from the dependence of the mean
spacing on the mean density of the star (e.g., Kjeldsen \& Bedding
1995):
 \begin{equation}
 \Delta\nu \propto M^{1/2}R^{-3/2}, 
 \label{eq:large1}
 \end{equation}
so that
 \begin{equation}
 \delta R / R =-2/3 \delta \Delta\nu / \Delta\nu.
 \label{eq:large2}
 \end{equation}
The absolute uncertainty in $\Delta\nu$ required to give a 1\,\%
uncertainty in $R$ is therefore given by:
 \begin{equation}
 \delta \Delta\nu_{1\,\%} = 0.01 \times 3/2 \Delta\nu = 0.015 \Delta\nu.
 \label{eq:large3}
 \end{equation}
The top right-hand panels of each figure again show estimates of the
large spacings, but plotted over a much larger range on the ordinate
to show poor estimates of the spacings.

The bottom left-hand panels of each figure show estimated spacings
extracted (by WJC) from analysis of autocorrelations of noise-free
power frequency spectra (see Figure~\ref{fig:borislim} and discussion
above). The different symbols show results for the frequency ranges
chosen by particular hounds. These results serve as a useful
``noise-free'' reference for the main (noisy timeseries) results shown
in the top panels of each figure.

Finally, the bottom right-hand panels show estimates of the
uncertainties on the large spacings returned by the hounds. The dashed
lines show standard deviations of the results of the hounds at each
apparent visual magnitude, $m_v$. The dotted lines again show the
uncertainties in $\Delta\nu$ required to give 1\,\% uncertainties on
the inferred stellar radii. The most striking aspect of the
uncertainty estimates is that they cover such a wide range.

 \section{Discussion}
 \label{sec:disc}

How should we interpret the precision in the results (i.e., the
observed scatter between hounds), and the uncertainty data plotted in
the bottom right-hand panel of each figure? In reaching a conclusion
we need to remember that all datasets on a given star had the same
mode realization noise, but different shot noise.

At the brighter magnitudes (lower noise spectra), scatter between
hounds is dominated by differences in implementation of the analysis
(what we might term \emph{reduction} noise), which includes the impact
of the different frequency ranges chosen by the hounds (see lower
left-hand panels of Figures~\ref{fig:panchores}
to~\ref{fig:katrinares}). The magnitude of this observed scatter is
larger than the characteristic internal precision on the estimated
spacings. By \emph{internal} precision, we mean the typical precision
we would expect to get were one of the hounds to analyse many datasets
having different shot \emph{and} mode realization noise. For example,
T.~Arentoft analyzed several fully independent timeseries realizations
of Boris (made by WJC), and found that the characteristic internal
precision on the estimated spacings was $\sim 0.04\,\rm \mu Hz$ at
$m_v=9$, and $\sim 0.20\,\rm \mu Hz$ at $m_v=11$. The observed scatter
between hounds is $\approx 0.5\,\rm \mu Hz$ at both magnitudes. At the
fainter magnitudes, where scatter between hounds increases, the
influence of different realizations of the shot noise comes into play,
and the internal precision also drops.

The quoted uncertainties should give robust estimates of the internal
precision. However, it is clear from Figures~\ref{fig:panchores}
to~\ref{fig:katrinares} that the quoted values cover a wide range at
each $m_v$. Some of the hounds' uncertainties were very much smaller
even that the expected internal precision. In several cases, this may
have been because those estimates came from standard least-squares
fits to autocorrelations of the power frequency spectra, where no
account had been taken of the strong correlations present in the
autocorrelation functions. (See also Chaplin et al. 2007b, for
discussion of a similar problem, using the cross correlation
function.)

 \section{Conclusion}
 \label{sec:conc}

The following conclusions may be drawn from the results:

\begin{enumerate}

 \item It is possible to extract robust estimates of the large
 frequency spacings, $\Delta\nu$, for many of the timeseries we
 made. Estimated spacings of the different hounds, returned at values
 of apparent visual magnitude where robust estimation is possible,
 show agreement at a level typically better than 1\,\%.

 \item The observed level of agreement implies a contribution to the
 uncertainty on inferred stellar radii of typically a few tenths of
 1\,\%.

 \item Scatter in the results increases significantly at $m_v=15$ for
 Pancho; at $m_v \ge 13$ for Boris; and at $m_v > 9$ for
 Katrina. Indeed, for the intrinsically fainter Katrina no results on
 the spacings were returned at $m_v \ge 11$.

 \item The observed scatter between hounds for the brighter stars
 tends to be larger than the characteristic internal precision
 expected, given the intrinsic quality of the data. The main
 contribution to this additional scatter appears to be reduction noise
 (i.e., differences in implementation of the analysis).

 \item Estimation of uncertainties in the large spacings, from fits
 made to prominent peaks in the autocorrelation of the power frequency
 spectrum, must make proper allowance for the strong correlations that
 are present in the autocorrelation function.

\end{enumerate}

The next phase of the exercise will involve true blind tests on single
timeseries of new stars. We then intend to use the results, together
with further tests on the data here, to inform development and
refinement by the KASC of the analysis pipeline for extracting the
frequency spacings. The results of the asteroFLAG exercises also need
to be understood in the context of the Kepler simulations undertaken
at Aarhus (see Christensen-Dalsgaard et al. 2007).

\acknowledgements

We are extremely grateful to the International Space Science Institute
(ISSI) for support provided by a workshop programme award. This work
was also supported by the European Helio- and Asteroseismology Network
(HELAS), a major international collaboration funded by the European
Commission's Sixth Framework Programme. We thank the VIRGO/SOHO team,
whose data we used. SOHO is a mission of international cooperation
between ESA and NASA.

\end{document}